\documentstyle[aps,prl]{revtex}

\begin{document}

\input epsf.sty
\twocolumn[\hsize\textwidth\columnwidth\hsize\csname %
@twocolumnfalse\endcsname

\draft

\widetext




\title{Instantaneous Spin Correlations in La$_2$CuO$_4$ }

\author{ R.\@ J.\@ Birgeneau, M.\@ Greven$^1$, M.\@ A.\@ Kastner,
Y.\@ S.\@ Lee, and B.\@ O.\@ Wells$^2$}

\address{Department of Physics, Massachusetts Institute of
Technology, Cambridge, MA 02139, USA}

\author{Y. \@ Endoh and K. \@ Yamada$^3$}

\address{Department of Physics, Tohoku University, Aramaki, Aoba-ku,
Sendai 980, 980-77 Japan}

\author{G.\@ Shirane}

\address{Brookhaven National Laboratory, Upton, NY 11973, USA}

\date{\today}
\maketitle

\begin{abstract} We have carried out a neutron scattering study of the
instantaneous spin-spin correlations in La$_2$CuO$_4$ (T$_N$ = 325~K)
over the temperature range 337~K to 824~K.  Incident neutron energies
varying from 14.7~meV to 115~meV have been employed in order to
guarantee that the energy integration is carried out properly.  The
results so-obtained for the spin correlation length as a function of
temperature when expressed in reduced units agree quantitatively both
with previous results for the two dimensional (2D) tetragonal material
Sr$_2$CuO$_2$Cl$_2$ and with quantum Monte Carlo results for the
nearest neighbor square lattice S=1/2 Heisenberg model.  All of the
experimental and numerical results for the correlation length are well
described without any adjustable parameters by the behavior predicted
for the quantum non-linear sigma model in the low temperature
renormalized classical regime.  The amplitude, on the other hand,
deviates subtly from the predicted low temperature behavior.  These
results are discussed in the context of recent theory for the 2D
quantum Heisenberg model.
\end{abstract}


\phantom{.}

]

\narrowtext

The physics of low dimensional quantum Heisenberg antiferromagnets has
been the subject of research ever since the advent of modern quantum
and statistical mechanics [1,2].  Interest in two dimensional (2D)
systems was heightened by the discovery of high temperature
superconductivity in the lamellar copper oxides [3].  Specifically, it
was realized early on that the parent compounds such as La$_2$CuO$_4$
correspond to rather good approximations to the S=1/2 2D
square-lattice quantum Heisenberg antiferromagnet (2DSLQHA)[4,5].  It
seems at least possible that the 2D magnetism may in some way be
essential to the superconductivity in the charge carrier doped
cuprates.  Further, the magnetism itself is of fundamental interest as
a quantum many body phenomenon in lower dimensions.

Early experiments by Endoh and co-workers [5] showed that over a wide
range of temperatures above the three dimensional N\'{e}el ordering
transition in La$_2$CuO$_{4+y}$ (that is, La$_2$CuO$_4$ with a small
amount of excess oxygen) the instantaneous spin-spin correlations were
purely two dimensional and that the correlation length diverged
exponentially in 1/T.  This led to a flurry of theoretical activity
[2] including most especially work based on the quantum non-linear
sigma model (QNL$\sigma$M) by Chakravarty, Halperin and Nelson (CHN)
[6] and Hasenfratz and Niedermayer (HN) [7]. These theories are all
based on the 2D Heisenberg Hamiltonian which for nearest neighbor
($nn$) interactions alone takes the form
\begin{equation} H =
J\sum_{<i,\delta_{nn}>}\vec{S}_i\cdot\vec{S}_{i+\delta_{nn}}
\end{equation} where the summation is over $nn$
pairs on a square lattice.

In La$_2$CuO$_4$, for temperatures below the tetragonal (I4/{\it mmm})
- orthorhombic ({\it Bmab}) structural phase transition temperature of
$T_{st}$ = 530~K, the leading terms in the spin Hamiltonian [8,9] are

\begin{eqnarray} H  & = J~  & (\sum_{<i,\delta_{nn}>}
\vec{S}_i\cdot\vec{S}_{i+\delta_{nn}}~+~\alpha_{nnn}\sum_{<i,
\delta_{nnn}>}\vec{S}_i \cdot \vec{S}_{i+\delta_{nnn}} \nonumber \\
& + & \alpha_{xy}\sum_{<i, \delta_{nn}>} S^c_i S^c_{i+\delta_{nn}}
+ \sum_{<i, \delta_{\perp j}>} \alpha_{\perp j} \
\vec{S}_i\cdot\vec{S}_{i+\delta_{\perp j}} \nonumber \\
& + & \alpha_{DM}\sum_{<i,\delta_{nn}>} ( - )^i \hat{a} \cdot\vec{S}_i \times
\vec{S}_{i+\delta_{nn}}).
\end{eqnarray}  Here, $\alpha_{nnn}$, $\alpha_{xy}$, $\alpha_{\perp j}$, and
$\alpha_{DM}$ represent the reduced next nearest neighbor in-plane
Heisenberg exchange coupling, XY anisotropy, interlayer coupling and
Dzyaloshinski-Moriya antisymmetric exchange, respectively, and $S^c_i$
is the {\bf c} component of the spin at site $i$. The fourth term in
Eq. (2) explicitly includes the two different out-of-plane neighbors
at $\delta_{\perp 1}$ and $\delta_{\perp 2}$.  Note that, as was
implicit in the work of Thio {\it et al.\@} [8], the sign of the
antisymmetric term changes on opposite sublattices because of the
opposite rotation of the CuO$_6$ octahedra.  This Dzyaloshinski-Moriya
term originates from a small rotation of the CuO$_6$ octahedra about
the $\hat a$ axis. In the tetragonal phase $\alpha_{DM} = 0$ and the
nearest neighbor out-of-plane effective coupling vanishes since
$\alpha_{\perp 1} = \alpha_{\perp 2}$.

The most complete experimental study to-date is on the material
Sr$_2$CuO$_2$Cl$_2$ [10] rather than La$_2$CuO$_4$.  The reasons for
this are twofold: First, Sr$_2$CuO$_2$Cl$_2$ is very difficult to dope
so that there are no complications arising from the effects of doped
electrons or holes on the spin correlations.  Second, since
Sr$_2$CuO$_2$Cl$_2$ is tetragonal down to the lowest temperatures
measured ($<$ 10~K), $\alpha_{DM}$=0 and the nearest neighbor
interplanar coupling vanishes to leading order, that is $\alpha_{\perp
1} = \alpha_{\perp 2}$.  As shown in Table (1) there is a small XY
anisotropy. In addition, from results in Sr$_2$Cu$_3$O$_4$Cl$_2$ [11],
we infer that there is a next nearest neighbor in-plane Heisenberg
exchange coupling which is about 8$\%$ of the nearest neighbor value.
To first order, the latter should simply lead to a slight
renormalization of the effective $J$ in Eq.\@ (1).  The XY anisotropy
will lead to a crossover from Heisenberg to XY behavior for
correlation lengths $\xi/a \stackrel{>}{\sim}$ 100.  Thus,
Sr$_2$CuO$_2$Cl$_2$ should be a good realization of the S=1/2 2DSLQHA
for length scales $\stackrel{<}{\sim}$ 100.  This has, in fact, been
confirmed in detail experimentally [10]; specifically, over a wide
range of length scales the 2D correlation length measured in
Sr$_2$CuO$_2$Cl$_2$ agrees quantitatively with results from quantum
Monte Carlo (QMC) calculations carried out on the Hamiltonian Eq.\@
(1) with S=1/2 [12-14].  The value for $J$ for Sr$_2$CuO$_2$Cl$_2$
listed in Table (1) is deduced from two magnon Raman scattering
measurements [15].

Both the QMC and the Sr$_2$CuO$_2$Cl$_2$ experimental results for the
correlation length in turn are quantitatively predicted by theory
based on the QNL$\sigma$M in the low temperature renormalized
classical (RC) regime [6, 7].  This comparison again involves no
adjustable parameters.  Surprisingly, this agreement holds for
correlation lengths as short as a few lattice constants.  This is far
outside of the temperature range where the QNL$\sigma$M-RC theory
should hold.  A plausible explanation for this unexpected agreement
has been given by Beard {\it et al.}[13]

In spite of the fact that the progenitor of this work was the
discovery of high temperature superconductivity in
La$_{2-x}$Ba${_x}$CuO$_4$ [3] together with the early work on the 2D
spin correlations in La$_2$CuO$_{4+y}$ [5], our \linebreak

\

\noindent
\begin{tabular}{|@{\hspace{.2in}}c@{\hspace{.2in}}|@{\hspace{.2in}}c@{\hspace{.2in}}|@{\hspace{.2in}}c@{\hspace{.2in}}|}  \hline
                & La$_2$CuO$_4$ & Sr$_2$CuO$_2$Cl$_2$   \\ \hline
S               & 1/2           & 1/2                   \\ \hline
T$_N$ (K)       & 325           & 256.5                 \\ \hline
$J$ (meV)       & 135           & 125                   \\ \hline
$\alpha_{nnn}$  & $\sim$ 0.08   & $\sim$ 0.08           \\ \hline
$\alpha_{DM}$   & $1.5 \times 10^{-2}$ & -              \\ \hline
$\alpha_{XY}$   & $-5.7 \times 10^{-4}$ & $-5.3 \times 10^{-4}$ \\ \hline
$\alpha_{\perp 1}-\alpha_{\perp 2}$ & $5 \times 10^{-5}$ & $\sim 10^{-8}$ \\ \hline
\end{tabular}
\begin{table}
\caption{N\'{e}el temperature, superexchange energy,
and corrections to the 2D Heisenberg Hamiltonian for La$_2$CuO$_4$ [9]
and Sr$_2$CuO$_2$Cl$_2$ [10].  $\alpha_{DM}$ and $\alpha_{XY}$ are
larger than the values quoted in references [9] and [10] by factors of
($Z_c/Z_g$) and $(Z_c / Z_g)^2$ respectively.  Here $Z_c$(1/2) $\simeq$
1.17 and $Z_g$(1/2) $\simeq$ 0.6 are the quantum renormalization
factors for the spin wave velocity and spin wave gap respectively.}
\end{table}

\noindent  knowledge of the spin correlations in stoichiometric
La$_2$CuO$_4$ is rather limited.  The primary correlation length data
for La$_2$CuO$_4$ originate from the neutron scattering study of
Keimer {\it et al.\@} [9] on a carrier-free single crystal of
La$_2$CuO$_4$ with T$_N$ = 325~K.  The Keimer {\it et al.\@} [9] data
on the correlation length and structure factor extend up to 550~K.
Their measurements are generally consistent with the
Sr$_2$CuO$_2$Cl$_2$, QMC and QNL$\sigma$M-RC results, but there appear
to be systematic discrepancies at the limit of the error bars for the
correlation length at both low and high temperatures.  These neutron
experiments were carried out using a single incident neutron energy of
31~meV.  It seems likely that the discrepancies are an experimental
artifact originating from the use of a single incident neutron energy
over a wide range of temperatures.  Alternatively, they could
represent a real effect originating from the antisymmetric exchange
and interplanar coupling terms in Eq.\@ (2) for La$_2$CuO$_4$.
Clearly, therefore, it is important to carry out a more complete study
of the spin-spin correlations in La$_2$CuO$_4$ in order to
characterize fully the magnetism in this parent compound of the
monolayer high temperature superconductors. Such data would also be
valuable for the interpretation of NQR results in $\rm La_2CuO_4$
[16].  Finally, there have been some important advances in our
understanding of the theory for the 2DSLQHA since the work of Greven
{\it et al.\@}[10] on Sr$_2$CuO$_2$Cl$_2$ and it is therefore of value
to re-examine the relationships between the results of experiments in
real systems and theory.

The experiments were carried out primarily on the H7 triple-axis
spectrometer at the High Flux Beam Reactor at Brookhaven National
Laboratory.  The measurements utilized the same single crystal of
La$_2$CuO$_4$ as employed by Keimer {\it et al.\@} [9]; this crystal
had a volume of about 1.5 cm$^3$.  Throughout this paper we use {\it
Bmab} orthorhombic axes; at T$_N$ =325~K the lattice constants are
$a=5.338~$\AA, $b=5.406~$\AA, and $c=13.141~$\AA.  We show in Fig.\@ 1
the temperature dependence of the (0 1 2) nuclear superlattice peak
intensity together with the reduced orthorhombic splitting (b-a)/(b+a)
[17].  As is evident from Fig.\@ 1, the sample of La$_2$CuO$_4$ shows
a sharp tetragonal-orthorhombic structural phase transition at
T$_{st}$ = 530.5$\pm$ 0.5~K.  The sharpness of the transition in turn
reflects the microscopic homogeneity of this sample.

The magnetic neutron scattering experiments were carried out in the
energy-integrating two axis mode.  For 2D systems the integration over
energy is carried out automatically in a two axis experiment provided
that the outgoing neutron wave vector $\vec k_f$ is perpendicular to
the 2D planes and provided that the neutron energy is significantly
larger than the characteristic energy $\omega_0$ of the spin
fluctuations at a given temperature [18].  From the theory of CHN [6,
19] one has
\begin{equation}
\omega^{CHN}_{0} = {c\over \xi}({T\over 2\pi\rho_s})^{1/2}
\end{equation}

\begin{figure}
\vspace{-7mm}
\centerline{\epsfxsize=3.0in\epsfbox{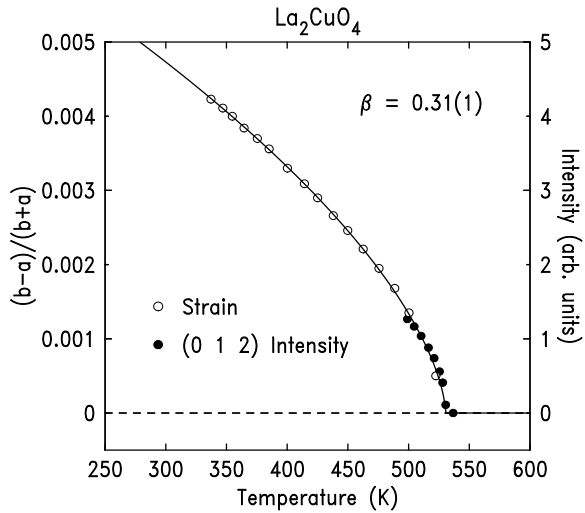}}
\vskip 5mm
\caption{
Orthorhombic splitting and $(0, 1, 2)$ superlattice peak intensity
versus temperature; the data are normalized relative to each other in
the temperature region of overlap.  The solid line is the result of a
fit to a power law A(T$_{st}$-T)$^{2\beta}$ with $\beta$ = 0.31 $\pm$
0.01 and T$_{st}$ = 530.5$\pm$0.5~K.
}
\label{Figure1}
\end{figure}

\noindent where $c$ and $\rho_s$ are the zero temperature spin
wave velocity and spin stiffness respectively.  For La$_2$CuO$_4$ this
becomes [20]
\begin{equation}
\omega^{CHN}_{0} \simeq \frac{850~{\rm meV\AA}}{\xi}
\sqrt{{T\over 1800}}
\end{equation}

Quantum Monte Carlo calculations by Makivi\'{c} and Jarrell [21] at
intermediate temperatures generally are well described by Eq.\@ (4)
but with an amplitude that is approximately twice as large.
Specifically, Makivi\'{c} and Jarrell [21] find that between 550~K and
800~K for $J=135$~meV, as in La$_2$CuO$_4$ [20], $\omega_{0}$ varies
from $\sim$ 25~meV to $\sim$ 63~meV.  Accordingly, the following
protocol was used in our measurements.  Neutrons with incident
energies E$_i$ = 14.7~meV were used at the lower temperatures.  With
increasing temperature and hence decreasing $\xi$ the incoming neutron
energy was progressively raised to 41~meV, 90~meV and 115~meV.  To
ensure that the energy integration was carried out correctly, it was
required that the results for the correlation length in the
temperature regions of overlap agreed with each other to well within
the experimental errors.

We show first in Fig.\@ 2 preparatory data taken at a temperature of T
= 328~K which is just above the 3D N\'{e}el temperature of 325~K.  The
incident neutron energy was E$_i$ = 3.6 meV which results in very high
momentum resolution.  The two peaks evident in Fig.\@ 2 orginate from
the two rods of scattering which are along $(1, 0, l)$ and $(0, 1,
l)$, respectively.  The equi-intensity of the two peaks implies that
at 328~K the 2D spin fluctuations have at least XY symmetry, that is,
at 328~K there is no measurable in-plane anisotropy induced by the
antisymmetric exchange terms in Eq.\@ (2).

\begin{figure}
\centerline{\epsfxsize=3.2in\epsfbox{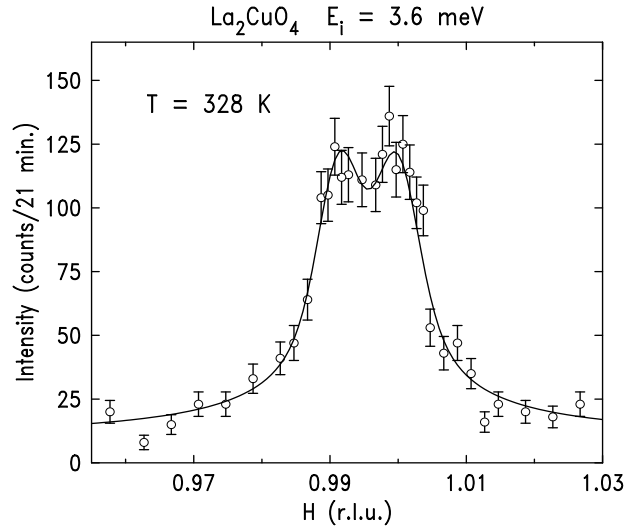}}
\vskip 5mm
\caption{
E = 3.6~meV two-axis scan across the 2D rods at $(1, 0, 1.59)$ and $(0,
1, 1.59)$.  The solid line is the result of a fit to two Lorentzians,
Eq. (5), centered about $(1, 0, l)$ and $(0, 1, l)$ respectively,
convolved with the instrumental resolution function.  The fit gives
$\xi^{-1}$ = 0.0011 $\pm$0.0004 reciprocal lattice units.
}
\label{Figure2}
\end{figure}

In Figs.\@ 3 and 4 some representative energy-integrating scans for
E$_i$ = 41 meV and E$_i$ = 115 meV are shown.  The collimations were
set to $20^{\prime}-10^{\prime}-$S$-10^{\prime}$ in both cases,
and for neutrons with E$_i$ = 41~meV a pyrolytic graphite filter was
used.  For E$_i$ = 115~meV, the experiment was carried out without a
filter in order to maximize the neutron flux.  Higher order
contamination from neutrons with energies above $\sim$ 400~meV is not
a concern as it results from the high-energy-tail of the thermal
neutron spectrum peaked at $\sim$ 30~meV.  The solid lines in Figs.\@
3 and 4 are the result of fits to the 2D Lorentzian form
\begin{equation} {\cal S}(\vec{q}_{2D}) = {{\cal S}(0)\over 1 +
q^2_{2D}\xi^2},
\end{equation}  where $\vec{q}_{2D}$ is the 2D deviation in wave
vector from the positions of the $(1, 0, l)$ and $(0, 1, l)$  rods,
convolved with the instrumental resolution function of the
spectrometer.

The results so-obtained for the inverse correlation length $\xi^{-1}$
are shown in Fig.\@ 5.  These data are consistent within the errors
with the earlier results of Keimer {\it et al.} [9], but are much more
precise and cover a wider range of temperatures.  The solid line is
the predicted behavior for the QNL$\sigma$M in the renormalized
classical regime [6, 7]; this will be discussed below.  The results
for the Lorentzian amplitude ${\cal S}(0)/\xi^2$ are shown in Fig.\@
6. The four sets of data are normalized to unity over the temperature
interval 450~K $\leq$ T $\leq$ 550~K.

We now compare the results in Fig.\@ 5 for the correlation length in
La$_2$CuO$_4$ with the predictions of various theories.  We begin with
the results of QMC calculations \linebreak

\begin{figure}
\vspace{-1mm}
\centerline{\epsfxsize=3.0in\epsfbox{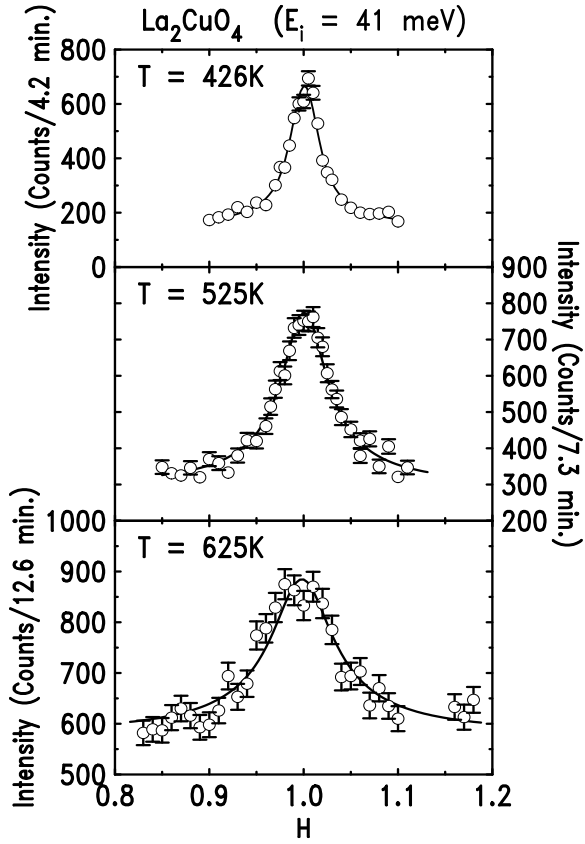}}
\vskip 5mm
\caption{
Representative energy-integrating two-axis scans in La$_2$CuO$_4$ with
E$_i$ = 41 meV and collimations 20$^\prime$ - 10$^\prime$ - $S$ -
10$^\prime$.  The solid lines are the result of fits to a 2D
Lorentzian scattering function Eq. (5) convolved with the resolution
function of the spectrometer.
}
\label{Figure3}
\end{figure}

\noindent  for Eq. (1) with $S = 1/2$.  Because
of both advances in computational techniques and the implementation of
finite-size scaling methods, QMC data now exist for $\xi$/a for the $S
= 1/2$ $nn$ 2DSLQHA for length scales varying from 1 to 350,000
lattice constants.  QMC results of Beard {\it et al.} [13] and Kim and
Troyer [14] are plotted in Fig.\@ 7 together with our experimental
results in La$_2$CuO$_4$.  The data are plotted in the reduced form
$\xi$/a vs.  $J$/T.  It is evident that the QMC and La$_2$CuO$_4$
results agree in absolute units over the complete temperature range
(337~K $<$ T $<$ 824~K) or equivalently, length scale range (3
$\stackrel{<}{\sim}$$ ~\xi$/a $\stackrel{<}{\sim}$ 115).  Thus, over
this range the 2D spin correlations in La$_2$CuO$_4$ are entirely
determined by the leading near-neighbor Heisenberg couplings and the
anisotropic in-plane plus interplanar terms in Eq.\@ (2) have no
measurable effect to within the uncertainty of our experiments.
Specifically, the tetragonal-orthorhombic structural phase transition
at 530~K does not manifest itself in the temperature dependence of the
correlation length.

We now consider the predictions of various analytical theories.  A
low-temperature theory for the 2DSLQHA \linebreak

\begin{figure}
\vspace{-9mm}
\centerline{\epsfxsize=3.0in\epsfbox{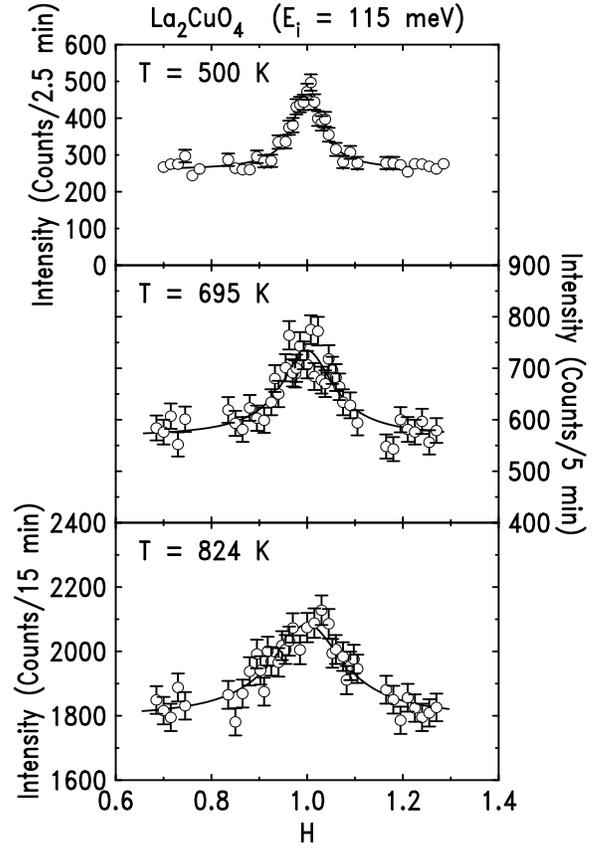}}
\vskip 8mm
\caption{
Representative energy-integrating two-axis scans in La$_2$CuO$_4$ with
E$_i$ = 115~meV and collimations 20$^\prime$ - 10$^\prime$ - $S$ -
10$^\prime$.  The solid lines are the results of fits to a 2D
Lorentzian scattering function convolved with the resolution function
of the spectrometer.
}
\label{Figure4}
\end{figure}

\noindent was formulated by Chakravarty,
Halperin, and Nelson [6], in which they obtained the static and
dynamic properties of the 2DSLQHA by mapping it onto the 2D quantum
non-linear $\sigma$ model.  The 2D QNL$\sigma$M is the simplest
continuum model which reproduces the correct spin-wave spectrum and
spin-wave interactions of the 2DSLQHA at long wavelengths and low
energies.  First, CHN argued that for $S \geq 1/2$ the $nn$ 2DSLQHA
corresponds to the region of the 2D QNL$\sigma$M in which the ground
state is ordered - the renormalized classical regime.  Then, CHN used
perturbative renormalization group arguments to derive an expression
for the correlation length to two-loop order, showing a leading
exponential divergence of $\xi$ versus inverse temperature.  Later,
Hasenfratz and Niedermayer [7] employed chiral perturbation theory to
calculate the correlation length more precisely to three-loop order.
In the RC scaling regime, the correlation length is given by

\begin{equation} {\xi\over a} = {e\over 8}{c/a\over
2\pi\rho_s}e^{2\pi\rho_s/T}
\left[ 1-{1\over2}\left({T\over 2\pi\rho_s}\right) + {\cal
O}\left({T\over2\pi\rho_s}\right)^2 \right],
\end{equation}

\begin{figure}
\vskip 5mm
\centerline{\epsfxsize=3.0in\epsfbox{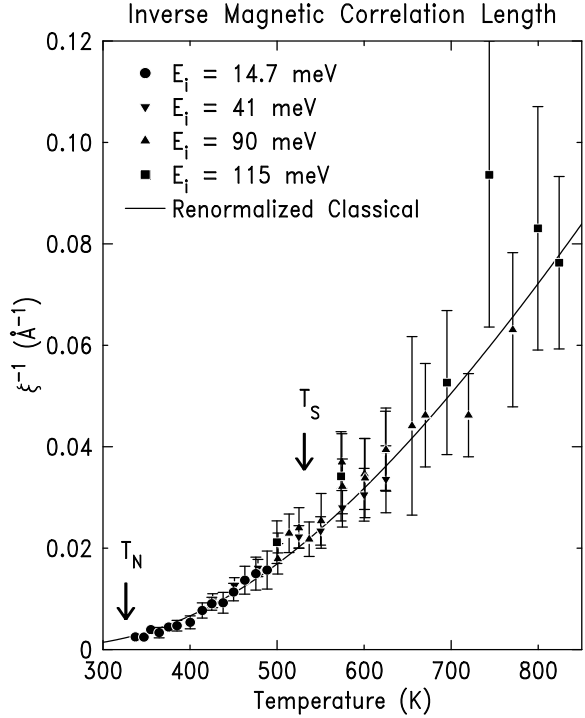}}
\vskip 5mm
\caption{
Inverse magnetic correlation length of La$_2$CuO$_4$.  The solid line
is Eq. (6) with $J = 135$ meV.  The N\'{e}el and structural transition
temperatures are indicated by arrows.
}
\label{Figure5}
\end{figure}

\noindent which we refer to as the CHN-HN formula.  The parameters $\rho_s$ and
$c$ are the macroscopic T = 0 spin stiffness and spin-wave velocity,
respectively.  For the nearest-neighbor 2DSLQHA, they are related to
the microscopic parameters $J, S$ and the lattice constant $a$
according to $\rho_s$ = $Z_\rho(S)S^2J$ and $c=Z_c(S)2\sqrt{2}aSJ$.
The coefficients $Z_\rho(S)$ and $Z_c(S)$ are quantum renormalization
factors, which can be calculated using spin-wave theory $(S\geq 1/2)$
[22, 23], series expansion $(S = 1/2, 1)$ [23], and Monte Carlo
techniques $(S= 1/2)$ [12-14].  For $S= 1/2$, the spin-wave
approximation gives $Z_\rho(1/2) \simeq 0.699$ and $Z_c(1/2)\simeq
1.18$ [22, 23].  The most precise values for $S = 1/2$ currently
available come from the QMC study of Beard {\it et al.} [13] who find
$c=1.657(2)Ja$, and $\rho_s = 0.1800(5)J$ and, for the T=0 sublattice
magnetization, $M_s = 0.30797(3)/a^2$.  These correspond to
$Z_c(1/2)=1.172$ and $Z_{\rho_s}(1/2)=0.72$.  The CHN-HN prediction
for the Lorentzian amplitude ${\cal S}(0)/\xi^2$ is
\begin{equation}  {{\cal S} (0)\over \xi^2} =
\mbox{A}2 \pi M^2_s\left({\mbox{T}\over2\pi\rho_s}\right)^2
\left[1+C_1 {\mbox{T}\over2\pi\rho_s}+{\cal
O}\left({\mbox{T}\over2\pi\rho_s}\right)^2\right],
\end{equation}

It is of interest to compare the QNL$\sigma$M predictions with the
corresponding predictions of the \underline{classical} spin model for
the $nn$ 2DSLHA at low temperatures [2,6,24,25]

\begin{figure}
\vskip 5mm
\centerline{\epsfxsize=3.0in\epsfbox{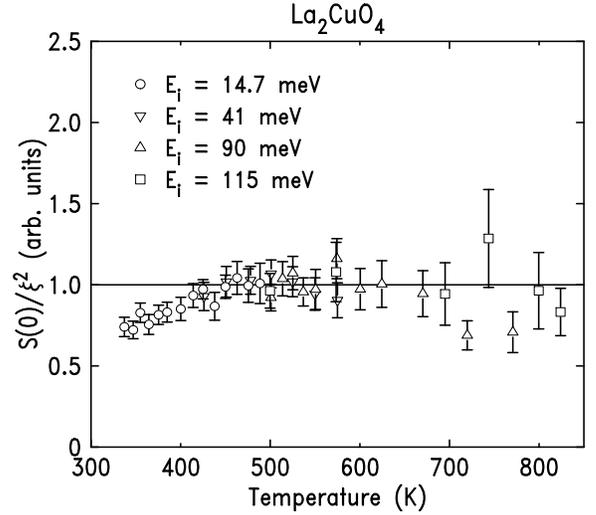}}
\vskip 5mm
\caption{
Lorentzian amplitude, $\cal{S}$(0)/$\xi^2$ versus temperature.  The
data for the different incident neutron energies are normalized to
unity in the temperature range 450~K $\stackrel{<}{\sim}$ T
$\stackrel{<}{\sim}$ 550~K.
}
\label{Figure6}
\end{figure}

\begin{figure}
\vskip 2mm
\centerline{\epsfxsize=3.0in\epsfbox{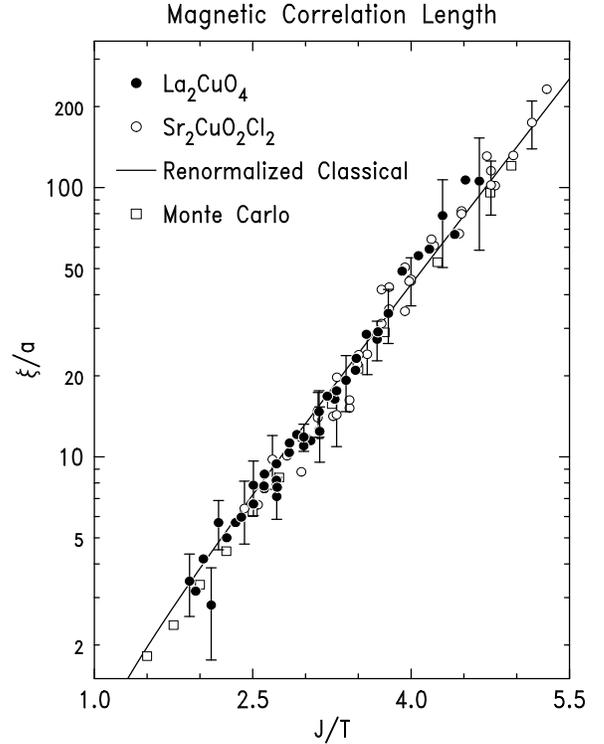}}
\vskip 5mm
\caption{
The logarithm of the reduced magnetic correlation length $\xi$/a
versus $J/T$.  The closed circles are data for La$_2$CuO$_4$ plotted
with $J = 135$ meV, the open circles are data for Sr$_2$CuO$_2$Cl$_2$
plotted with $J = 125$ meV [10], and the open squares are the results
of the Monte Carlo computer simulations [12-14].  The solid line is
the theoretical prediction without adjustable parameters of the
2DQNL$\sigma$M for the renormalized classical regime Eq. (6).
}
\label{Figure7}
\end{figure}

\begin{equation}
{\xi\over a} = 0.0125 {\mbox{T}\over2\pi\rho_{cl}}
e^{2\pi\rho_{cl}/\mbox{T}}
\left[1-b_1{\mbox{T}\over2\pi\rho_{cl}} + {\cal
O}\left({\mbox{T}\over\rho_{cl}}\right)^2\right]
\end{equation} and
\begin{equation} {{\cal S} (0)\over \xi^2} = \mbox{A}
2\pi M^2_s \left({\mbox{T}\over
2\pi\rho_{cl}}\right)^2\left[1+\mbox{C}_1
{\mbox{T}\over2\pi\rho_{cl}}+{\cal O}
\left({\mbox{T}\over\rho_{cl}}\right)^2\right]
\end{equation} where for classical unit vector spins, $\rho_{cl}=J$.  For
large $S$ quantum spins, one finds that the classical limit is
approached smoothly as a function of $S$ provided that temperature is
measured in units of $JS(S+1)$, implying one that should take
$\rho_{cl}=JS(S+1)$ [2].  This choice for $S = 1/2$, gives $\rho_{cl}
= 0.75J$ compared with $\rho_s = 0.18J$.  The arguments in the
exponentials in Eq.\@ (6) and (8) then differ by more than a factor of
4 -- a very dramatic difference between renormalized classical and
classical scaling behavior.

An alternative theoretical analysis of the 2DSLQHA has been carried
out by Cuccoli {\it et al.} [26] in which they treat quantum
fluctuations in a self-consistent Gaussian approximation separately
from the classical contribution.  In their approach, which they label
the purely-quantum self-consistent harmonic approximation (PQSCHA),
the quantum spin Hamiltonian is rewritten as an effective classical
Hamiltonian, where the temperature scale is renormalized due to
quantum fluctuations, and the new classical spin length appears as
$S+1/2$.  Defining the reduced temperature as t= T/$\{J(S+1/2)^2\}$,
the correlation length for the 2DSLQHA is then simply given by
\begin{equation}
\xi(\mbox{t}) =
\xi_{cl}(\mbox{t}_{cl}) \quad\mbox{with}\quad \mbox{t}_{cl} =
{\mbox{t}\over\theta^4(\mbox{t})}.
\end{equation}
Here, $\xi_{cl}$ is the correlation length for the corresponding
classical 2D square-lattice nn Heisenberg model, which is given by
Eq.\@ (8) at low temperatures and may be obtained from classical spin
Monte Carlo calculations at higher temperatures [27], and
$\theta^4$(t) is a temperature renormalization parameter.  The PQSCHA
is most accurate in the limit where the quantum fluctuations are weak,
and correspondingly $\theta^4$ is near unity.  This is the case for
large spin, for example, S=5/2, and indeed this has recently been
studied experimentally by Leheny {\it et al.} [28] who have measured
${\cal S}$($\vec{q}_{2D}$) in the S = 5/2 2DSLQHA material
Rb$_2$MnF$_4$. They follow a field-temperature trajectory which
approaches the bicritical point in the phase diagram and which
accordingly should show pure 2D Heisenberg behavior.  In this case the
PQSCHA predicts the correlation length precisely with no adjustable
parameters over the inverse temperature range 0.5 $< \rho_{cl}/T <$ 2
or equivalently, the length scale range 1$<\xi/a
\stackrel{<}{\sim}$ 100.  We note from Eq.\@ (8) and (9) that for the
classical model T always appears scaled by $\rho_{cl}$.  Thus the
quantum effects in the PQSCHA can be thought of simply as a
temperature dependent renormalization of $\rho_{cl}$, that is,
$\rho_{cl} \rightarrow
\theta^4$(t)$\rho_{cl}$.

Finally, for the QNL$\sigma$M there may be a crossover from
renormalized classical to quantum critical behavior with increasing
temperature [6].  In the QC regime heuristically one expects
\begin{equation}
\xi /a = 0.8 {c/a\over T-T_{QC}}
\end{equation}
with T$_{QC}\geq$ 0 adjustable [2,6,10]. We emphasize that this
anticipated crossover is a property of the QNL$\sigma$M and it may or
may not occur for quantum spins on a 2D lattice.

The solid line in Fig.\@ 5 is the QNL$\sigma$M-RC prediction, Eq.\@
(6) with $c$ and $\rho_s$ from Beard {\it et al.} [13] As observed
previously for Sr$_2$CuO$_2$Cl$_2$ [10], Eq.\@ (6) describes the
measured correlation length of $\rm La_2CuO_4$ extremely well without
adjustable parameters over the temperature range 337~K $<$ T $<$ 824~K,
or equivalently, the length scale range, $\sim 3
\stackrel{<}{\sim}\xi$/a$\stackrel{<}{\sim} 110$.  All of the data for
$\xi$/a from each of quantum Monte Carlo, Sr$_2$CuO$_2$Cl$_2$ and
La$_2$CuO$_4$ together with Eq. (6) are plotted in the universal form
$\xi$/a vs. $J$/T in Fig.\@ 7.  The evident universal behavior is, of
course, both pleasing and reassuring.  The good agreement of all of
the experimental and numerical results with the low temperature
QNL$\sigma$M-RC predictions down to very small values of $J$/T at
first appears to be quite puzzling.  The QMC study, Ref. [13],
suggests that this agreement is, at least in part, accidental.
Specifically, in crossing over from the low temperature continuum
QNL$\sigma$M to the discrete lattice S=1/2 Heisenberg model the higher
order terms in Eq.\@ (6) conspire such that over the measured
temperature range the deviation of $\xi$/a from Eq.\@ (6) is never
more than 15$\%$ which is well within the experimental error.

We now focus on the high temperature behavior in La$_2$CuO$_4$.  We
show in Fig.\@ 8 the La$_2$CuO$_4$ correlation length data together
with the predictions from QNL$\sigma$M-RC (Eq.\@ (6)), QNL$\sigma$M-QC
Eq.\@ (11), high temperature series expansion [29] and the PQSCHA
which involves Eq. (10) combined with results of classical Monte Carlo
simulations [27].  As observed previously for Sr$_2$CuO$_2$Cl$_2$ [10]
as well as for both S = 1/2 2DSLQHA QMC calculations [14] and high
temperature series expansion results [29],the QNL$\sigma$M-QC
prediction, Eq.\@ (11), disagrees strongly with the experimental
results in La$_2$CuO$_4$.  This is, perhaps, not surprising given the
extremely short length scales at the relevant temperatures.
Specifically, at these short distances, the continuum QNL$\sigma$M
approach which underlies the possible QC behavior is probably no
longer valid.  By contrast the PQSCHA which corresponds to classical
scaling for the pure 2D \linebreak

\begin{figure}
\vspace{-5mm}
\centerline{\epsfxsize=3.0in\epsfbox{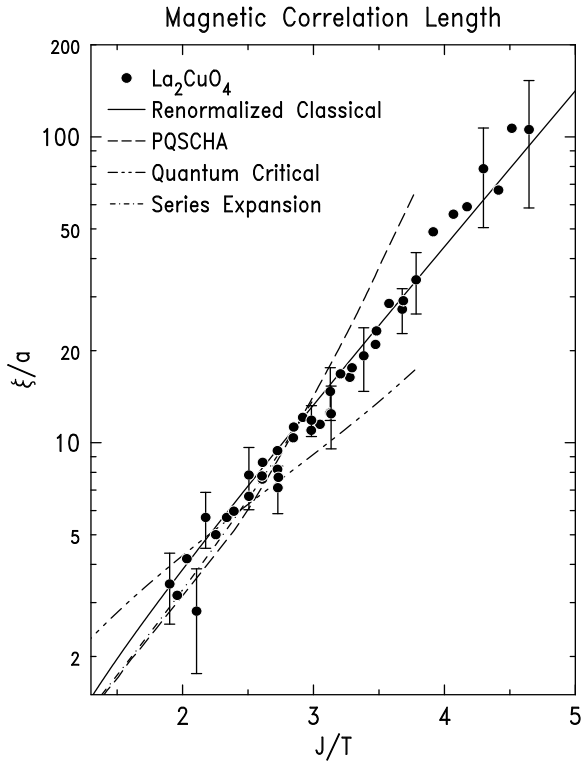}}
\vskip 5mm
\caption{
The logarithm of the reduced magnetic correlation length $\xi$/a
versus $J/T$ compared with the predictions of various theories
including renormalized classical behavior, Eq. (6), quantum critical
behavior, Eq. (11), the PQSCHA, Eq. (10), and high temperature series
expansion [29].
}
\label{Figure8}
\end{figure}

\noindent Heisenberg  model [2,30] agrees reasonably well
in absolute units with no adjustable parameters for length scales up
to about $\xi$/a $\sim$ 15.  As expected, the PQSCHA breaks down at
lower temperatures and larger length scales.  Thus, if there is a
crossover in the correlation length, it is from classical scaling to
renormalized classical scaling with decreasing temperature.  Clearly,
it is very important that theory for this crossover from the high
temperature PQSCHA classical scaling regime to the low temperature
QNL$\sigma$M-RC regime be developed.

Finally, we discuss the behavior of the structure factor ${\cal
S}(0)$.  The leading divergence of ${\cal S}(0)$ is determined by
$\xi^2$.  This is confirmed by the results for ${\cal S}(0)$ in
La$_2$CuO$_4$ displayed in Fig.\@ 6 which shows that ${\cal
S}(0)$/$\xi^2$ is approximately constant over the complete temperature
range.  This is equivalent to the statement that for the 2D S=1/2 QHA
the critical exponent $\eta_2=0$.  We note that in 3D $\eta_3 \simeq
0.04$, [31] that is, ${\cal S}(0) \sim \xi^{1.96}$ whereas for the 1D
S=1/2 QHA one has the remarkable result that ${\cal S}(0)
\sim (ln \xi)^{3/2}$ [32], which implies $\eta_1=2$.

The temperature dependent correction factors (c.f. Eq.\@ (7)) beyond
the leading $\xi^2$ divergence are problematic.  Specifically, Greven
{\it et al.\@} [10] find in their measurements in Sr$_2$CuO$_2$Cl$_2$
that over the length scale 5$\stackrel{<}{\sim}
\xi$/a$\stackrel{<}{\sim}$200, ${\cal S}(0)/\xi^2$ is independent of
temperature to within the errors.  By contrast, QMC [12-14] and high
temperature series expansion [29,30] studies of the S=1/2 nn 2DSLQHA
find ${\cal S}(0)/\xi^2 \sim$ T$^2$ over about the same range of
length scales.  In the S = 5/2 2DSLQHA material $\rm Rb_2MnF_4$,
Leheny {\it et al.} [28] find a clear crossover at $\xi$/a$\sim$4 from
${\cal S}(0)/\xi^2\sim$T$^2$ behavior to a much weaker dependence of
${\cal S}(0)/\xi^2$ on T.  The data for ${\cal S}(0)/\xi^2$ in
La$_2$CuO$_4$ shown in Fig.\@ 5 are clearly inconsistent with T$^2$
behavior over the complete temperature range but would allow a gradual
crossover as found in Rb$_2$MnF$_4$.  This lack of universality in
${\cal S}(0)/\xi^2$ seems surprising given the robust universality of
the behavior for $\xi$/a (Fig.\@ 7).  Of course, departures from the
low temperature QNL$\sigma$M-RC behavior may occur at different
temperatures for different quantities. It is also possible that the
terms in Eq. (2) beyond the nearest neighbor Heisenberg coupling will
effect ${\cal S}(0)/\xi^2$ more than they effect $\xi$ itself.

In summary, we have carried out a neutron scattering study of the
instantaneous spin-spin correlations in La$_2$CuO$_4$ (T$_N$=325~K)
over the temperature range 337~K to 824~K.  Incident neutron energies
varying from 14.7~meV to 115~meV have been employed in order to
guarantee that the energy integration is carried out properly.  The
results so-obtained for the spin correlation length as a function of
temperature when expressed in reduced units agree quantitatively both
with previous results for the 2D tetragonal material
Sr$_2$CuO$_2$Cl$_2$ and with quantum Monte Carlo results for the
nearest neighbor square lattice S = 1/2 Heisenberg model.  All of the
experimental results for the correlation length are well described
without any adjustable parameters by the behavior predicted for the
quantum non-linear sigma model in the low temperature renormalized
classical regime.  The structure factor, on the other hand, deviates
subtly from the predicted low temperature behavior although the
leading $\xi^2$ behavior is confirmed.  The correlation length data at
high temperature agree reasonably well with predictions of the PQSCHA
which corresponds to classical scaling with quantum corrections for
the 2D Heisenberg model.  We therefore hypothesize that in
La$_2$CuO$_4$ there is a gradual crossover from renormalized classical
to classical scaling with increasing temperature.
\

\begin{center}{\bf Acknowledgements}
\end{center}

We would like to acknowledge helpful discussions and communications on
these results with S. Charkravarty, T. Imai, R. Leheny, R.R.P. Singh,
V. Tognetti, P. Verrucchi, U.-J.  Wiese and J. Zinn-Justin.  This work
was supported in part by a Grant-In-Aid for
Scientific Research from
the Japanese Ministry of Education, Sc ience, Sports and Culture, by a
Grant for the Promotion of Science from the Science and Technology
Agency and by CREST.  Work at Brookhaven National Laboratory was
carried out under Contract No. DE-AC02-98-CH10886, Division of
Material Science, U.S. Department of Energy.  The research at MIT was
supported by the National Science Foundation under Grant No.
DMR97-04532 and by the MRSEC Program of the National Science
Foundation under Award No. DMR98-08941.

\

\begin{center}{\bf References}
\end{center}

$^1${\it Permanent address:} Dept. of Applied Physics and Stanford
Synchrotron Radiation Laboratory, Stanford University, Stanford, CA
94305, USA

$^2${\it Permanent address:} Dept. of Physics, Brookhaven National
Laboratory, Upton, NY 11973, USA

$^3${\it Permanent address:} Institute for Chemical Research, Kyoto
University, Uji 611-0011, Japan

\begin{enumerate}

\item H. A. Bethe, Z. Phys. {\bf 71}, 205 (1931).

\item For reviews see:  E. Manousakis, Rev. Mod.
Phys. {\bf 63}, 1 (1991); N. Elstner, Int. J. Mod. Phys.  {\bf B11}
1753 (1997).

\item J. G. Bednorz and K. A. M$ \rm \ddot{u}$ller, Z.
Phys.  {\bf B64}, 189 (1986).

\item P. W. Anderson, Science {\bf 235}, 1196 (1987).

\item Y. Endoh, K. Yamada, R. J. Birgeneau, D. R.
Gabbe, H. P. Jennsen, M. A. Kastner, C. J. Peters, P.  J. Picone, T.
R. Thurston, J. M. Tranquada, G.  Shirane, Y. Hidaka, M. Oda, Y
Enomoto, M. Suzuki, and T. Murakami, Phys. Rev.  {\bf B37}, 7443
(1988).

\item S. Chakravarty, B.I. Halperin, and D.R. Nelson,
Phys. Rev.  {\bf B39}, 2344 (1989).  For a detailed discussion of the
quantum critical region see also A. Chubukov, S. Sachdev and Y. Ye,
Phys. Rev.  {\bf B49} 2344 (1994).

\item P. Hasenfratz and F. Niedermayer, Phys. Lett. B
{\bf 268}, 231 (1991).

\item T. Thio, T. R. Thurston, N. W. Preyer, P. J.
Picone, M. A. Kastner, H. P. Jennsen, D. R. Gabbe, C.  Y. Chen, R. J.
Birgeneau, and A. Aharony, Phys. Rev.  B {\bf 38}, 905 (1988).

\item B. Keimer, N. Belk, R. J. Birgeneau, A.
Cassanho, C. Y. Chen, M. Greven, M. A. Kastner, A.  Aharony, Y. Endoh,
R. W. Erwin, and G. Shirane, Phys.  Rev.  {\bf B46}, 14034 (1992).

\item M. Greven, R. J. Birgeneau, Y. Endoh, M. A.
Kastner, M. Matsuda, and G. Shirane, Z. Phys. B {\bf 96}, 465 (1995).

\item F. C. Chou, A. Aharony, R. J. Birgeneau, O.
Entin-Wohlman, M. Greven, A. B. Harnig, M. A.  Kastner, Y.-J. Kim, D.
S. Kleinberg, Y. S. Lee, and Q.  Zhu, Phys. Rev. Lett. {\bf 78}, 535
(1998); Y.-J. Kim (private communication).

\item M. Makivi\'{c} and H.-Q. Ding, Phys. Rev.
{\bf B43}, 3562 (1991).

\item B. Beard, R. J. Birgeneau, M. Greven, and U.-J.
Wiese, Phys. Rev. Lett {\bf 80}, 1742 (1998).

\item J.-K. Kim and M. Troyer, Phys. Rev. Lett.
{\bf 80}, 2705 (1998).

\item Y. Tokura, S. Koshihara, T. Arima, H. Takagi,
S. Ishibashi, T. Ido, and S. Uchida, Phys. Rev.  {\bf B41}, 11657
(1990).

\item T. Imai, C. P. Schlicter, K. Yoshimura, M.
Katoh, and K. Kosuge, Phys. Rev. Lett. {\bf 71}, 1254 (1993).

\item R. J. Birgeneau, C. Y. Chen, D. R. Gabbe, H.
P. Jennsen, M. A. Kastner, P. J. Picone, T. Thio, T.  R. Thurston, and
H. L. Tuller, Phys. Rev. Lett. {\bf 59}, 1329 (1987).

\item R. J. Birgeneau, J. Skalyo, Jr., and G. Shirane,
Phys. Rev.  {\bf B3}, 1736 (1971).

\item S. Ty$\rm \breve{c}$, B. I. Halperin, and S.
Chakravarty, Phys. Rev. Lett. {\bf 62}, 835 (1989).

\item G. Aeppli, S. M. Hayden, H. A. Mook, Z. Fisk,
S. W. Cheong, D. Rytz, J. P. Remeika, G. P. Espinosa, and A. S.
Cooper, Phys. Rev. Lett. {\bf 62}, 2052 (1989).

\item M. Makivi\'c and M. Jarrell, Phys. Rev. Lett.
{\bf 68}, 1770 (1992).

\item J. Igarashi, Phys. Rev.  {\bf B46}, 10763 (1992).

\item R.R.P. Singh, Phys. Rev.  {\bf B39}, 9760 (1989); C. Hamer, Z.
Weihong, and J. Oitmaa, Phys. Rev. B {\bf 50}, 6877 (1994).

\item Th. Jolicoeur and J. C. LeGouillou, Mod. Phys.
Lett. {\bf B5}, 593 (1991).

\item D. R. Nelson and R. A. Pelcovits, Phys. Rev.
{\bf B16}, 2191 (1977).

\item A. Cuccoli, V. Tognetti, R. Vaia, and P. Verrucchi, Phys. Rev.
{\bf B56}, 14456 (1997). 

\item J.-K. Kim, Phys. Rev. {\bf D50}, 4663 (1994).

\item R. L. Leheny, R. J. Christianson, R. J.
Birgeneau, and R. W. Erwin, Phys.\@ Rev.\@ Lett.\@ {\bf 82}, 418
(1999).

\item A. Sokol, R.L. Glenister and R.R.P. Singh, Phys.\@ Rev.\@ Lett.\@, {\bf
72}, 1549 (1994).

\item N. Elstner, A. Sokol, R. R. P. Singh, M. Greven,
and R.J. Birgeneau, Phys.\@ Rev.\@ Lett.\@ {\bf 75}, 938 (1995).

\item See for example L.C. LeGuillon and J. Zinn-Justin, Phys. Rev.  {\bf B21},
3976 (1980).

\item O.A. Starykh, A.W. Sandvik, and R.R.P. Singh, Phys. Rev.  {\bf B55},
14953 (1997).

\end{enumerate}

\end{document}